\documentclass[prb,twocolumn,showpacs,citeautoscript,amsmath]{revtex4}

\usepackage{graphicx}
\setkeys{Gin}{width=6.6 cm}   

\begin{document}

\title{Listening to the coefficient of restitution and the 
gravitational acceleration of a bouncing ball}

\author{C. E. Aguiar} \email{carlos@if.ufrj.br}
\author{F. Laudares}  \email{f_laudares@hotmail.com}
\affiliation{Instituto de F{\'\i}sica, Universidade Federal do Rio de Janeiro \\
Cx.P.~68528, Rio de Janeiro, 21945-970, RJ, Brasil}

\begin{abstract}
We show that a well known method for measuring the coefficient of restitution
of a bouncing ball can also be used to obtain the gravitational acceleration.
\end{abstract}


\maketitle

Three contributions to this journal have described how to measure the coefficient 
of restitution between a ball and a flat surface using the sound made by 
the collision of the ball with the surface \cite{Bernstein,Smith,Stensgaard}. 
The procedure reported in these papers is to drop the ball vertically on a horizontal 
surface, allow it to bounce several times, while recording the sound produced by the impacts. 
Analysis of the recording gives the time intervals between successive rebounds, and
from these the coefficient of restitution is obtained.

The evolution of the techniques described in these papers is a nice example 
of how the development of microcomputers has changed the science teaching laboratory.
In 1977, Bernstein \cite{Bernstein} detected the sound with a microphone, 
amplified and filtered the signal, and fed it to a pen recorder.
Smith, Spencer and Jones \cite{Smith}, in 1981, connected the microphone to 
a microcomputer via a homemade data collection and interface circuit, and then uploaded the
resulting data to a larger computer for analysis and graphical display.
In 2001, Stensgaard and L{\ae}gsgaard \cite{Stensgaard} used the microphone input 
of a PC sound card to make the recording, reducing the experimental equipment to 
basically a standard microcomputer.

To see how the coefficient of restitution $\epsilon$ is related to the time 
between bounces, note that if $\epsilon$ is constant (independent of velocity), 
and air resistance is negligible, the velocity of the ball just after the 
$n$th bounce on the fixed surface is given by
\begin{equation}
   v_n = v_0 \epsilon^n 
\label{vn}
\end{equation}
where $v_0$ is the velocity just before the first impact. 
The time-of-flight $T_n$ between the $n$th and $(n+1)$th collisions is 
proportional to $v_n$,
\begin{equation}
   T_n = \frac{2 v_n}{g} \; ,  \quad n=1,2,\dots   
\label{Tv}
\end{equation}
where $g$ is the gravitational acceleration. Thus 
\begin{equation}
   T_n = T_0 \epsilon^n \;,
\label{Tn}
\end{equation}
where we have defined $T_0 \equiv 2 v_0 / g$ . 
Taking the logarithm of both sides of Eq.~(\ref{Tn}) we obtain
\begin{equation}
   \log T_n =  n \log \epsilon + \log T_0 \; ,
\label{logTn}
\end{equation}
so that the plot of $\log T_n$ vs $n$ is a straight line of slope 
$\log\epsilon$ and intercept $\log T_0$.
Thus, as long as it is independent of velocity, the coefficient of restitution
can be obtained by fitting the straight-line of Eq.~(\ref{logTn}) 
to the time-of-flight data.

\begin{figure}[b]
\begin{center}
\includegraphics{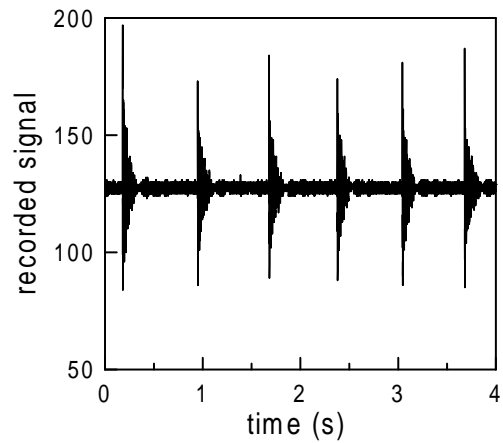}
\end{center}
\caption{
The sound of a ball bouncing on a horizontal surface. 
The zero sound level corresponds to 128 in the vertical axis.}
\label{signal}
\end{figure}

The purpose of this note is to point out that this straight-line fit can also be 
used to determine another physical quantity of interest: the gravitational acceleration $g$.
The (rather simple) observation is that, if the ball is released from a known 
height $h$, then $T_0=(8h/g)^{1 / 2}$, and
\begin{equation}
   g = \frac{8h}{{T_0}^2} \; .
\label{g}
\end{equation}
Thus, just as the slope parameter of Eq.~(\ref{logTn}) fixes the coefficient of
restitution, the intercept parameter determines the acceleration of gravity 
(if the easily measured initial height $h$ is known). 

In order to check how this works in practice, we have dropped a ``superball'' from 
a measured height onto a smooth stone surface and recorded the sound produced
by the successive impacts. 
The recording was made with the microphone and sound card of a PC
running Windows, using the sound recorder program that comes with the
operating system. 
The sampling frequency was 22,050~Hz, resulting in a time resolution of 45~$\mu$s. 
The audio file, stored in the binary WAV format, was converted to ASCII text format
with the shareware program \textsc{awave~audio} \cite{awave}. 
The recorded signal is plotted in Fig.~\ref{signal}, where the pulses 
corresponding to individual impacts are easily recognized (only the first 
six collisions are shown). 
We have used 8-bit resolution in the recording, so that data values can only go 
from 0 to 255.
The no-signal value corresponds to 128.

\begin{figure}[t]
\begin{center}
\includegraphics{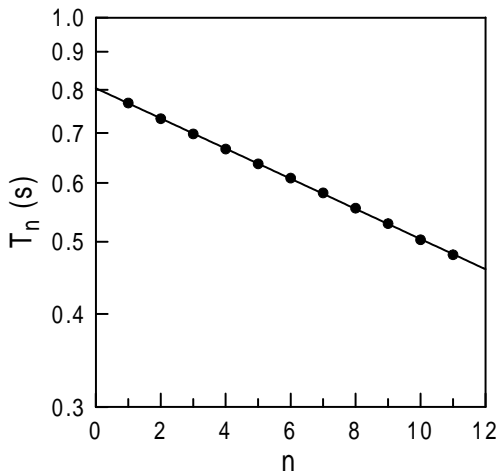}
\end{center}
\caption{Time-of-flight $T_n$ between impacts $n$ and $n+1$. 
The line is the least-squares fit using Eq.~(\ref{logTn}).}
\label{tof}
\end{figure}

The time intervals $T_n$ between collisions $n$ and $n+1$ were obtained directly 
through inspection of the ASCII sound file. 
They are plotted in Fig.~\ref{tof} (in logarithmic scale) as a function of $n$. 
The least-squares fit of the $T_n$ data set to Eq.~(\ref{logTn})  
gives 
\begin{subequations}
\label{fitpars}
\begin {eqnarray}
  && \epsilon = 0.9544 \pm 0.0002      \; , 
\\
  && T_0 = 0.804 \pm 0.001 \mbox{\ s}  \; . 
\end{eqnarray}
\end{subequations}
The best-fit line is also shown in Fig.~\ref{tof}.

The ball was released from a height $h = 79.4 \pm 0.1$~cm above the surface.
Taking this and the adjusted $T_0$ into Eq.~(\ref{g}), we obtain for the 
gravitational acceleration 
\[ g=982 \pm 3 \mbox{\ cm/s}^2 \;. \] 
For comparison, the value of $g$ in Rio de Janeiro is 978.8~cm/s$^2$.

\begin{figure}[t]
\begin{center}
\includegraphics{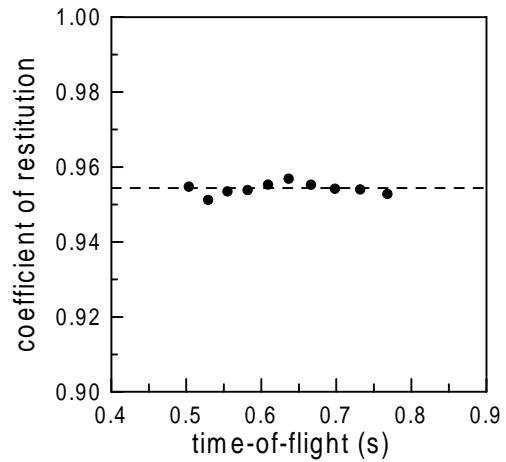}
\end{center}
\caption{The coefficient of restitution $\epsilon = T_{n+1}/T_n$ as a function of
the time-of-flight $T_n$, for the data of Fig.~\ref{tof}.
The dashed line indicates the adjusted value given in Eq.~(\ref{fitpars}).}
\label{epsT}
\end{figure}

The applicability of the method described above depends on $\epsilon$ being 
constant over the range of impact velocities involved in the experiment. 
That this condition is satisfied in the present case is seen in Fig.~\ref{epsT}, 
where the coefficient of restitution for an impact at velocity $v_n$, 
$\epsilon = v_{n+1}/v_n=T_{n+1}/T_n$, is plotted as a function of $T_n$
(recall that $v_n \propto T_n$, see Eq.~(\ref{Tv})).
The coefficients of restitution for the different impacts are all very close to 
the least-squares value given in Eq.~(\ref{fitpars}), indicated by the dashed line 
in Fig.~\ref{epsT}.

\begin{figure}[b]
\begin{center}
\includegraphics{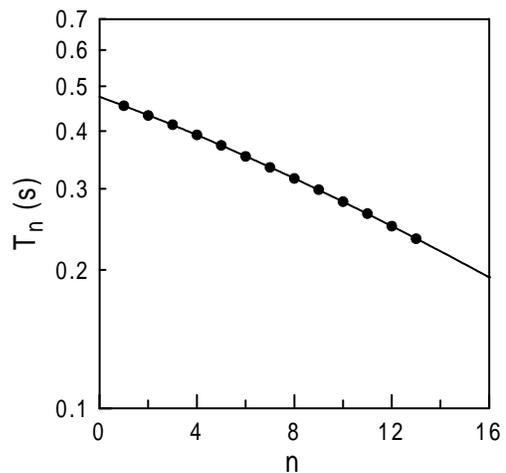}
\end{center}
\caption{Time-of-flight $T_n$ between impacts $n$ and $n+1$. 
The curve is the least-squares fit using Eq.~(\ref{Tn2}).}
\label{tof2}
\end{figure}

\begin{figure}[t]
\begin{center}
\includegraphics{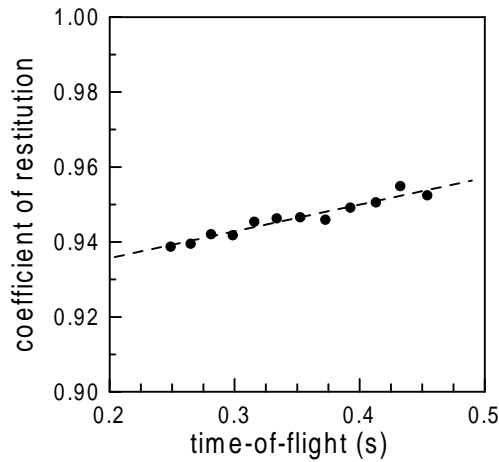}
\end{center}
\caption{The coefficient of restitution $\epsilon = T_{n+1}/T_n$ as a function 
of $T_n$, for the data of Fig.~\ref{tof2}. 
The dashed line is the linear relation of Eq.~(\ref{linear}) 
with the adjusted parameters given in Eq.~(\ref{fitpars2}).}
\label{epsT2}
\end{figure}

A case in which the coefficient of restitution depends on the velocity is shown in
Fig.~\ref{tof2}, where we display the times of flight of a superball dropped from
$h = 27.5 \pm 0.1$~cm onto a wood surface.
A plot of $\epsilon$ at each collision, shown in Fig.~\ref{epsT2}, reveals 
a clear dependence of the coefficient of restitution on the time-of-flight
(or impact velocity).
Assuming a linear relation between $\epsilon$ and $T$, as suggested by 
Fig.~\ref{epsT2},
\begin{equation}
   \epsilon = \epsilon_0 (1 + \alpha  T) \;,
\label{linear}
\end{equation}
we obtain an extension of Eq.~(\ref{Tn})
\begin{equation}
   T_n = T_0 {\epsilon_0}^n  \prod_{i=0}^{n-1}(1 + \alpha  T_i) \; .
\label{Tn2}
\end{equation}

The least-squares fit of Eq.~(\ref{Tn2}) to the data shown in Fig.~\ref{tof2} 
gives
\begin{subequations}
\label{fitpars2}
\begin{eqnarray}
   && \epsilon_0 = 0.921 \pm 0.001   \; , 
\\
   && \alpha = 0.078 \pm 0.003 \mbox{\ s}^{-1}  \; , 
\\
   && T_0 = 0.4752 \pm 0.0005 \mbox{\ s}  \; . 
\end{eqnarray}
\end{subequations}
The curves corresponding to these parameters are also shown in 
Figs.~\ref{tof2} and \ref{epsT2}.
The above value for $T_0$ yields 
\[ 
    g = 974 \pm 5 \mbox{\ cm/s}^2 \;,
\]
again a very reasonable value.
Consideration of the velocity dependence of the coefficient
of restitution was important in order to get an accurate result; 
had we assumed a constant $\epsilon$, we would obtain 
$g = 935 \pm 10$~cm/s$^2$. 

To summarize, we have seen that the value of the gravitational acceleration 
is a useful by-product of experiments devised to ``hear'' the coefficient of 
restitution of a bouncing ball. 
The measurement of $g$ is particularly simple if  the coefficient of restitution 
is independent of the impact velocity, but more complicated cases can also be 
handled.

\ 

After this work was completed we learned of a recent paper by Cavalcante \emph{et al.}
\cite{Cavalcante}, in which $g$ was measured using the sound of a bouncing ball. 
The analysis presented in the paper is, however, somewhat different from ours. 
Another related reference is the article by Guercio and Zanetti \cite{Guercio}
in this journal.


\end{document}